\documentclass[12pt]{article}
\usepackage[dvips]{color}
\usepackage{graphicx}
\usepackage{amsthm,amssymb,amsmath,graphics,times}

\newtheorem{theorem}{Theorem}

\theoremstyle{definition}

\theoremstyle{remark}

\newcommand{\NaTimeConstant}{\alpha_{ _{\rm Na}}}
\newcommand{\KTimeConstant}{\alpha_{ _{\rm K}}}
\newcommand{\CaTimeConstant}{\alpha_{ _{\rm Cx}}}
\newcommand{\GTimeConstant}{\alpha_{ _{\rm G}}}

\newcommand{\Kabattery}{E_{\rm K}}
\newcommand{\Naabattery}{E_{\rm Na}}
\newcommand{\Caabattery}{E_{\rm Cx}}
\newcommand{\Gabattery}{E_{\rm G}}

\newcommand{\Kbarconductance}{{\bar g}_{ _{\rm K}}}
\newcommand{\Nabarconductance}{{\bar g}_{ _{\rm Na}}}
\newcommand{\Cabarconductance}{{\bar g}_{ _{\rm Cx}}}
\newcommand{\Gbarconductance}{{\bar g}_{ _{\rm G}}}

\newcommand{\Kzero}{Q_{\rm K}}
\newcommand{\Nazero}{Q_{\rm Na}}
\newcommand{\Cazero}{Q_{\rm Cx}}
\newcommand{\Gzero}{Q_{\rm G}}

\newcommand{\Keta}{\eta_{ _{\rm K}}}
\newcommand{\Naeta}{\eta_{ _{\rm Na}}}
\newcommand{\Caeta}{\eta_{ _{\rm Cx}}}
\newcommand{\Geta}{\eta_{ _{\rm G}}}

\newcommand{\Kphi}{\phi_{ _{\rm K}}}
\newcommand{\Naphi}{\phi_{ _{\rm Na}}}
\newcommand{\Caphi}{\phi_{ _{\rm Cx}}}
\newcommand{\Gphi}{\phi_{ _{\rm G}}}

\newcommand{\Vvoltage}{V}
\newcommand{\Vvoltageprime}{{V}'}

\newcommand{\Kspopen}{{\epsilon}_{ _{\rm K}}}
\newcommand{\Naspopen}{{\epsilon}_{ _{\rm Na}}}
\newcommand{\Caspopen}{{\epsilon}_{ _{\rm Cx}}}
\newcommand{\Gspopen}{{\epsilon}_{ _{\rm G}}}

\begin{document}

%\centerline{\Large \sffamily Information Rate in Golden Ratio Distribution}

%\centerline{\Large \sffamily for A Neural Communication System}

%\centerline{\Large \sffamily Golden Ratio Information for Neural Communication}

\centerline{\Large \sffamily Golden Ratio Information for Neural Spike Code}

%\centerline{\Large \sffamily }

\bigskip

\centerline{Bo Deng, Department of Mathematics,
University of Nebraska-Lincoln, Lincoln, NE 68588}

%\centerline{\sffamily Bo Deng\footnote{Department of Mathematics,
%University of Nebraska-Lincoln, Lincoln, NE 68588. Email: {\tt
%bdeng@math.unl.edu}}}

\bigskip
\noindent \textbf{Abstract: Spike bursting is a ubiquitous
feature of all neuronal systems. Assuming
the spiking states form an alphabet for a communication system,
what is the optimal information precessing rate? and what is the channel
capacity? Here we demonstrate that the quaternary alphabet of spike number code
gives the maximal processing rate, and
that a binary source
in Golden Ratio distribution gives rise to the channel capacity.
A multi-time scaled neural circuit is shown to satisfy
the hypotheses of this neural communication system. }

\bigskip
\noindent\textbf{1. Introduction.}
Claude E. Shannon's mathematical theory for communication
(\cite{Shan48}) laid a corner stone for the information age in the mid-20th
century. In his model, there is an information source to produce
messages in sequences of a source alphabet, an encoder/transmitter
to translate the messages into signals in time-series of a
channel alphabet, a channel to transmit the signals, and
a receiver/decoder to convert the signals back to the
original messages. The
mapping from the source alphabet to the channel alphabet is
a source-to-channel encoding. Any communication
system is characterized by two intrinsic performance parameters.
One, its \textit{mean transmission rate} (MTR) (in bit per time) for all
possible information sources, such as your Internet connection
speed for all kinds of sources, both wanted and unwanted. Two, its
\textit{channel capacity} (CC) which a particular information source or a
particular source-to-channel encoding scheme can take advantage to
transmit at the fastest data rate.

One property that is ubiquitous to all neurons is their
generation of electrical pulses across the cell membrane.
The form can be in voltage or any type of ionic current.
The form is not important to our discussion but only the state
of excitation or quiescence. We
further simplify our discussion by considering only those type
of neurons or neuronal circuit models which are capable of
spike bursts. A spike burst is a segment of time-series which
both begins and ends with a quiescent phase but with a sequence
of fast oscillations of equal amplitude and similar periods
in between. In terms
of dynamics, the quiescent phase
or the refractory phase is slow, and the spiking phase
is fast, typical of fast-slow, multi-timed neural models,
c.f. \cite{Rinz85,Term92,Guck97,Erme10}.
The fast-slow distinction between the two phases
and the spike number count allow us to
transform the analogue wave form into
a discrete information system as an obvious
assumption, see also \cite{Perk68}.

So, if we construct a communication system
using one neuron as an encoder/transmitter and another
as a receiver/decoder, then what would
be the best MTR and CC? More aptly, in a network of neurons,
any interneuron plays the role of transmission channel, the
same questions arise. That is, how much information
the neuron can process in a unit time when it is open for any
information and when it is restricted
to information of a particular type?

%%%%%%%%%%%%%%%%%%%%%%%%%%%%%%%%%%%%%%%%%%%%%%%%%%%%%%%%%%%%%%%%%%%%%%
%%%%%%%%%%%%%%%%%%%%%%%%%%%%%%%%%%%%%%%%%%%%%%%%%%%%%%%%%%%%%%%%%%%%%%
%%%%%%%%%%%%%%%%%%%%%%%%%%%%%%%%%%%%%%%%%%%%%%%%%%%%%%%%%%%%%%%%%%%%%%
\begin{table}%[ht]
\begin{center}
\caption{Definitions for Communication System
Parameters}\label{tabDefitions}
\begin{tabular}{rcl}
\noalign{\medskip} \hline \noalign{\medskip}
${\mathcal A}_n=\{b_1,b_2,\dots,b_n\}$ &---& System alphabet for encoding, transmitting, and decoding.\\
$\tau=\{\tau_1,\tau_2,\dots,\tau_n\}$ &---& Base
transmitting/processing time $\tau_k$ for base
$b_k$.\\
$p=\{p_1,p_2,\dots,p_n\}$ &---& Probability distribution
over ${\mathcal A}_n$ for an
encoded source.\\
$H(p)=\sum_{k=1}^n p_k\log_2(1/{p_k})$ &---&
Averaged information entropy in bit per symbol of\\
& & a particular source with encoded distribution $p$.\\
$T(p,\tau)=\sum_{k=1}^n p_k\tau_k$ &---&
Averaged transmitting/processing time per symbol of\\
& & a particular source with encoded distribution $p$.\\
$R(p,\tau)=H(p)/T(p,\tau)$ &---&
Particular transmission rate in bit per time of\\
& & a particular source with encoded distribution $p$.\\
$H_n=\log_2 n$ &---& Maximal entropy $H_n=\max_p\{H(p)\}$ with the\\
& & equiprobability, $p_k=1/n,\ 1\le k\le n$, for all sources.\\
$R_n(\tau)=H_n/\sum_{k=1}^n\tau_k/n$ &---&
Mean transmission rate (MTR) in bit per time for all\\
& &   sources with the equiprobability.\\
$K_n(\tau)=\max_p\{R(p,\tau)\}$ &---& Channel capacity (CC)
with $p_k=p_1^{\tau_k/\tau_1},\ \sum_{k=1}^n p_1^{\tau_k/\tau_1}=1$. \\
\noalign{\smallskip} \hline
\end{tabular}
\end{center}
\end{table}
%%%%%%%%%%%%%%%%%%%%%%%%%%%%%%%%%%%%%%%%%%%%%%%%%%%%%%%%%%%%%
%%%%%%%%%%%%%%%%%%%%%%%%%%%%%%%%%%%%%%%%%%%%%%%%%%%%%%%%%%%%%
%%%%%%%%%%%%%%%%%%%%%%%%%%%%%%%%%%%%%%%%%%%%%%%%%%%%%%%%%%%%%

%\bigskip\noindent\textbf{4. Optimal Quadrary Code.}
%From a practical point of view, a communication system is only
%able to use a finite system alphabet ${\mathcal A}_n$ (see Table
%\ref{tabDefitions}) to encode a
\bigskip
\noindent\textbf{2. Measurements of Communication System.}
Table \ref{tabDefitions} lists the definitions of
essential parameters and variables for any
information system (\cite{Reza94}). Given a source whose messages are
sequences of its source alphabet ${\mathcal
S}_\ell=\{s_1,s_2,\dots,s_\ell\}$, a source-to-channel encoding is
a mapping from the source alphabet ${\mathcal S}_\ell$ to the set
of finite sequences of the system alphabet ${\mathcal A}_n$ so
that a source sequence in ${\mathcal S}_\ell$ is translated to a
sequence in ${\mathcal A}_n$. A particular source
when source-to-channel coded in ${\mathcal A}_n$ is characterized
by a probability distribution $p$, with
$p_k$ being the probability to find
letter $b_k$ at any position of the encoded messages. According to
the information theory, letter or base $b_k$ contains $\log_2(1/p_k)$ bit of
information for the encoded source, and a typical base on average
contains $H(p)$ bit of information for the encoded source. In
statistical mechanics, the quantity entropy, $H(p)$, is a measure
of the disorder of a system, and in information theory, it
is the measure of diversity for the encoded source.
It is a simple and basic fact that the maximal entropy is reached
if and only if the probability distribution is
the equiprobability, $p_k=1/n,\ 1\le k\le n$, and the maximal
entropy is $H_{n}=\log_2(n)$.

From the designer's point view of a communication system,
the system is not for a { particular}
source with a particular encoded distribution, but rather for
{all} possible sources. For example, the Internet is not
designed for a particular source but rather for all sources, such
as texts, videos, talk-show radios, and so on.
In other words, it is only reasonable to assume that
each base $b_k$ will be used by some source, and it will be used
equally likely when averaged over all sources. Hence,
for any $n$-alphabet communication system, the maximal
information entropy $H_{n}=\log_2(n)$ in bit per base is a
system-wise payoff measure for the alphabet.

This does not mean that the larger the alphabet size the better
the system. The balance lies in the consideration of the payoff
against cost. The primary cost is the times the system takes to
process the alphabet, such as to represent or to transmit the
bases. Assume a base $b_k$ takes a fixed amount
of time, $\tau_k$, to process. Then for the
equiprobability distribution, the average processing time is
$T_n(\tau)=\sum_{k=1}^n\tau_k/n$ in time per base. Hence, the key
performance parameter for an $n$-alphabet communication system is
this payoff-to-cost ratio, $R_n(\tau)=H_n/T_n(\tau)$ in bit per
time, referred to as the {\em mean transmission rate} (MTR) or the {\em
transmission rate} for short. This is an intrinsic measure for all
communication systems, regardless of the size nor the nature of
their alphabets. As a result, different systems can be objectively
compared. In the example of Internet, the transmission rate is the
measure we use to compare different means of connection, such as
coaxial cable, optic fiber, or satellite. Notice also that the
mean transmission rate is determined by all sources,
and it is in this sense that the transmission
rate is a passive measurement of a communication system.

From the perspective of a particular source (user), its particular
payoff-to-cost ratio is the {\em source transmission rate},
$R(p,\tau)=H(p)/T(p,\tau)$, with $p$ being its distribution
over the system alphabet ${\mathcal A}_n$. This rate may be
faster or slower
than the mean rate $R_n(\tau)$. In other words, there is a
potential gain for a particular source to exploit
the system so that its bit
rate $R(p,\tau)$ is no worse than the mean. The
mathematical problem is to maximize the source transmission rate
$R(p,\tau)$ over all choices of the distribution $p$.
Solution to the optimization problem gives rise to the
{\em channel capacity} (CC), denoted by $K_n$.

\bigskip
\noindent\textbf{3. Result for Neural Spike Code.}
The number of spikes in a burst for a neuron is called the \
spike number and without introducing extra notation we denote it
by $b_k$ (as letters for the alphabet ${\mathcal A}_n$)
with the subscript $k\ge 1$ for the spike number of the burst.
The resultant alphabet system is referred to as the
neural spike code.

We consider first the simplest case when the process time
for the code progresses like the natural number:
$\tau_1=a,\tau_2=2a,\dots, \tau_k=ka$, for code base
$b_1, b_2,\dots, b_k$, respectively, for a fixed
parameter $a$. That is, the 1-spike base takes 1 unit
of time in $a$ to process, the 2-spike base takes 2 units
of time in $2a$ to process and so on. Without loss of
generality we can drop $a$ by assuming $a=1$ for now.

For the MTR, $R_n=H_n/[\sum_{k=1}^n k/n]$, of $\mathcal A_n$, we have
\[
R_n(\tau)={\log_2n}/{(n+2)/2}
\]
since $\sum_{k=1}^n k/n=n(n+1)/2/n$. For $n=2,3,\dots, 12$,
the values of $R_n$ are
\[
0.67,\ \    0.79,\ \    0.80,\ \    0.77,\ \    0.74,\ \    0.70,\ \
0.67,\ \    0.63,\ \    0.60,\ \    0.58,\ \    0.55
\]
That is, the optimal spike code is $\mathcal A_4$ with the
MTR $R_4=0.80$, as one
can easily show the function $\log_2 x/(x+2)/2$ is a decreasing
function for $x>4$.

For the CC, $K_n(\tau)$, consider the binary spike case
$\mathcal A_2$ first. To simplify notation, we use $p=p_1$
and $q=p_2$, $p+q=1$. Then, $H=-p\log_2p-q\log_2q$, $T=p\tau_1+q\tau_2
=p+2q$, and $R(p,q)=H/T$. We use the Lagrange multiplier method
to the find the maximum of $R(p,q)$ subject to the constraint
$g(p,q)=p+q=1$. This is to solve the following system of equations
\[
\begin{array}{l}
 R_p=\lambda g_p \\
 R_p=\lambda g_p \\
 g=1
\end{array}
\quad\hbox{ $\Longrightarrow$ }\quad
  \begin{array}{l}
  [(-\log_2p-1/{\rm ln}2)T-H]/T^2 =\lambda \\
  \left[(-\log_2q-1/{\rm ln}2)T-2H\right]/T^2 =\lambda \\
  p+q=1
  \end{array}
\]
Equate the left-sides of the first two equations
and simplify to get
\[
T\log_1p=T\log_2q+H
\]
implying that $\log_2p-\log_2q=H/T=R$. That is, at the maximal
distribution $(p,q)$, the maximal rate $K_2$ is
\begin{equation}\label{optR1}
R=H/T=\log_2(p/q)\ \Longleftrightarrow\ \log_2q=\log_2p-H/T.
\end{equation}
To find $(p,q)$, we use the relation $q=1-p$ from the
constraint $g(p,q)=1$ and the identity above to rewrite
$T=p+2q=2-p$ and then to replace all $q$ in $H$ as follows
\[
  \begin{array}{l}
  H= -p\log_2p-q\log_2q\\
  \quad = -p\log_2p-(1-p)[\log_2p-H/T]\\
  \quad = -p\log_2p-(1-p)\log_2p+(1-p)H/T\\
  \quad = -\log_2p+(T-1)H/T\\
  \quad = -\log_2p+H-H/T\\
 \end{array}
\]
which is simplified to
\begin{equation}\label{optR2}
R=H/T=-\log_2p.
\end{equation}
From Eq.(\ref{optR1}) and  Eq.(\ref{optR2}) we have
\[
-\log_2p=\log_2(p/q) \ \Longleftrightarrow\ 1/p=p/q \ \Longleftrightarrow\ (p+q)/p=p/q.
\]
The last equality shows $p/q$ is the Golden Ratio with
\[
p=\Phi=\frac{\sqrt 5 -1}{2}=0.6180\hbox{ and } q=p^2=\Phi^2
\]
and the channel capacity is
\[
K_2=R=-\log_2\Phi=0.6943>R_2=0.67,
\]
better than the binary MTR.

In fact, this result is a special case of the following theorem
which is a variation of Shannon's result from \cite{Shan48}.
A proof is a straightforward generalization of the Golden Ratio
case above.
\begin{theorem}\label{thmCapacity} For an $n$-alphabet communication
system, its source transmission rate $R(p,\tau)=H(p)/T(p,\tau)$
reaches a unique maximum $K_n(\tau)=\max_p\{R(p,\tau)\}$ at an
encoded source distribution $p$ which is the solution to the
following equations,
\begin{equation}\label{eqCCapacity}
p_k=p_1^{\tau_k/\tau_1}\hbox{\ \ for \ \ $1\le k\le n$\ \ and\ \ }
\sum_{k=1}^n p_1^{\tau_k/\tau_1}=1,
\end{equation}
and the maximal rate (the channel capacity) is
$K_n(\tau)=-\log_2 p_1/\tau_1$.
\end{theorem}

\begin{proof} (For review only.)
Since the maximization is independent from the base
presenting time $\tau$, we will drop all references of it from the
function $T$ and $R$. The proof is based on the Lagrange
multiplier method to maximize $R(p)$ subject to the constraint
$g(p)=\sum_{k=1}^np_k=1$. This is to solve the joint equations:
$\nabla R(p)=\lambda\nabla g(p), g(p)=1$, where $\nabla$ is the
gradient operator with respect to $p$ and $\lambda$ is the
Lagrange multiplier. Denote $R_{p_k}=\partial R/\partial p_k$.
Then the first system of equations becomes
$R_{p_k}=[H_{p_k}T-H\tau_{p_k}]/T^2=\lambda g_{p_k}=\lambda$,
componentwise. Write out the partial derivatives of $H$ and $T$
and simplify, we have $-(\log_2 p_k+1/\ln 2)T-H\tau_k=\lambda T^2$
for $k=1,2,\dots, n$. Subtract the equation for $k=1$ from each of
the remaining $n-1$ equations to eliminate the multiplier
$\lambda$ and to get a set of $n-1$ new equations: $-(\log_2
p_k-\log_2 p_1)T-H(\tau_k-\tau_1)=0$ which solves to
\begin{equation*}%\label{eqR}
R=[\log_2(p_k/p_1)]/(\tau_1-\tau_k)=\log_2[(p_k/p_1)^{1/(\tau_1-\tau_k)}].
\end{equation*}
Introducing a new quantity $\eta:=2^R=2^{H/T}$ or $H=T\log_2\eta$
we can rewrite the equation above as $\eta=(p_k/p_1)^{1/(\tau_1-\tau_k)}$
and equivalently
\begin{equation}\label{eqR}
p_k=\eta^{\tau_1-\tau_k}p_1
\end{equation}
for all $k$. Next we express the entropy $H$ in terms of
$\eta$ and $p_1, \tau_1$, substituting out all $p_k$:
\begin{equation*}
\begin{split}
H&=-\sum_{k=1}^np_k\log_2
p_k=-\sum_{k=1}^np_k[(\tau_1-\tau_k)\log_2\eta+\log_2 p_1]\\
&=-[\tau_1\log_2\eta-\sum_{k=1}^np_k\tau_k\log_2\eta+\log_2
p_1]=-[\tau_1\log_2\eta+\log_2 p_1]+T\log_2\eta,
\end{split}
\end{equation*}
where we have used $\sum_{k=1}^np_k=1$ and
$T=\sum_{k=1}^np_k\tau_k$. Since we have by definition
$H=T\log_2\eta$, cancelling $H$ from both sides of the equation
above gives
$\log_2 p_1+\tau_1\log_2\eta =0$ and consequently
\begin{equation}\label{eqR2}
R=\log_2\eta =\log_2 [p_1^{-1/\tau_1}]\hbox{ or } \eta=p_1^{-1/\tau_1}
\end{equation}
and from (\ref{eqR})
\begin{equation}\label{eqR3}
p_k=\eta^{\tau_1-\tau_k}p_1=p_1^{\tau_k/\tau_1}
\end{equation}
Last, solve the equation $f(p_1)=g(p)=\sum_{k=1}^np_1^{\tau_k/\tau_1}=1$ for
$p_1$. Since $f(p_1)$ is strictly increasing in $p_1$ and
$f(0)=0<1$ and $f(1)=n>1$, there is a unique solution $p_1\in
(0,1)$ so that $f(p_1)=1$. By (\ref{eqR2}) and (\ref{eqR3}),
the channel capacity is $K_n=R(p)=-[\log_2
p_1]/\tau_1 = -[\log_2 p_k]/\tau_k$. This completes the
proof.
\end{proof}

The Golden Ratio case is a corollary to the theorem with
the assumption that $\tau_2=2\tau_1$ for $n=2$. And for the
special case when $\tau_k=k\tau_1$ with $n>2$, the CC distribution
satisfies $p_k=p_1^k$ for $1\le k\le n$ with
\begin{equation}\label{eqGenGold}
p_1+p_1^2+p_1^3+\cdots +p_1^n=1.
\end{equation}
Denote the solution by $p_1=\beta_n$ for $\mathcal A_n$,
then we can easily show that
$\beta_n$ is a decreasing sequence in $n$, converging to $0.5$ and
bounded exactly from above by the Golden ratio $\beta_2=\Phi=0.6180$.

\bigskip
\noindent\textbf{4. Multi-time Scaled Neuron Model.} In \cite{Deng19},
a neuron model was discovered that remains symmetric under the
conductance and resistance transformation: $r=1/g$ and
$g=1/r$. The symmetry gives rise to the conductance
characteristics for both ion channels and protein channels:
\begin{equation}\label{eqGVcharact22}
\phi_X(V_X,\eta_X,Q_X)={\rm heaviside}({\rm sign}(\eta_X)(V_X-Q_X))\tanh^2[{|\eta_X|}(V_X-Q_X)/{2}].
\end{equation}
The model looks exactly the same under the transformation $gr=1$ and $\psi=1/\phi$,
the resistance characteristics,
leading to a unique model for any given neuron, rather than innumerable {\em ad hoc}
models as conventionally is the case.
This model then predicts that the phenomenon of spontaneous firing
of individual ion channels (\cite{Zago88,Sakm13}) by ways of quantum tunneling
(\cite{Nawa22}) is both sufficient and necessary,
marking the first transition from the quantum realm to the microscopic world
in neuronal modeling. The model automatically gives rise to
different time scales for ion and protein channels,
permitting dramatic simplifications in dimensional reduction.
The model also clearly lays out a blueprint for
circuit implementation by the channel characteristics (\ref{eqGVcharact22}).
The model is capable of both action potential propagation and spike-burst
generation. Here we consider a four-channel neuron model:
\begin{equation}\label{eqCRmodel}
\left\{\!\!\!\!\begin{array}{ll} & C\Vvoltageprime=
-[\Kbarconductance n (\Vvoltage-\Kabattery)+\Nabarconductance m (\Vvoltage-\Naabattery)\\
& \hskip 1in +\Gbarconductance h (\Vvoltage-\Gabattery)+\Cabarconductance c (\Vvoltage-\Caabattery)]\\
& {n}' = \displaystyle \KTimeConstant\sqrt{(n+\Kspopen)/({\Kphi}(V)+\Kspopen)}(\Kphi(V)-n)\\
& {m}' = \displaystyle \NaTimeConstant\sqrt{(m+\Naspopen)/{(\Naphi}(V)+\Naspopen)}(\Naphi(V)-m)\\
& {h}' = \displaystyle \GTimeConstant\sqrt{(h+\Gspopen)/{(\Gphi}(V)+\Gspopen)}(\Gphi(V)-h)\\
& {c}' = \displaystyle \sigma \CaTimeConstant\sqrt{(c+\Caspopen)/{(\Caphi}(V)+\Caspopen)}(\Caphi(V)-c)\\
\end{array}\right.
\end{equation}
where Na is for the sodium channel, K is for the potassium channel, G is for the sodium-potassium
gating channel, and the fourth channel, Cx, can be a calcium channel, or a chlorine channel,
or a protein channel, which are needed for spike-burst. Parameter $\sigma$ takes only a
fixed sign value, $+1$ or $-1$, because both can generate spike burst. Without the
spontaneous firing parameters $\epsilon_{\rm X}$, the model encounters a dividing-zero singularity
and a flat-lining equilibrium, meaning the neuron is neither functional nor alive.
All results outlined above are obtained in \cite{Deng19}.

The rate parameters, $\alpha_{ _{\rm X}}$, naturally make the model
multi-time scaled. For example, the full 5-dimensional system (\ref{eqCRmodel})
can be reduced to a 3-dimensional system below by assuming the gating
and the sodium channels to be the fastest with $\NaTimeConstant, \GTimeConstant$
sufficiently large so that $m=\Naphi(V)$ and $h=\Gphi(V)$:
\begin{equation}\label{eq3DSpikeModel}
\left\{\!\!\!\!\begin{array}{ll} & C\Vvoltageprime=-[\Kbarconductance n (\Vvoltage-\Kabattery)
+\Nabarconductance \Naphi(V) (\Vvoltage-\Naabattery)\\
& \hskip 1in +\Gbarconductance \Gphi(V) (\Vvoltage-\Gabattery)+\Cabarconductance c (\Vvoltage-\Caabattery)]\\
& {n}' = \displaystyle \KTimeConstant\sqrt{(n+\Kspopen)/({\Kphi}(V)+\Kspopen)}(\Kphi(V)-n)\\
& {c}' = \displaystyle \sigma \CaTimeConstant\sqrt{(c+\Caspopen)/{(\Caphi}(V)+\Caspopen)}(\Caphi(V)-c)\\
\end{array}\right.
\end{equation}
Simulations can be down on both with their dynamics indistinguishable for
large $\NaTimeConstant, \GTimeConstant$. System (\ref{eq3DSpikeModel}) is
another multi-time system, with the $V$-equation fast, the $n$-equation slow
or comparable, and the $c$-equation slower. The lower dimensional reduction
(\ref{eq3DSpikeModel}) can be advantageous for analytical manipulations
(c.f. \cite{Chen23}).

%%%%%%%%%%%%%%%%%%%%%%%%%%%%%%%%%%%%%%%%%%%%%%%%%%%%%%%%%%%%%
%%%%%%%%%%%%%%%%%%%%%%%%%%%%%%%%%%%%%%%%%%%%%%%%%%%%%%%%%%%%%
%%%%%%%%%%%%%%%%%%%%%%%%%%%%%%%%%%%%%%%%%%%%%%%%%%%%%%%%%%%%%
\begin{figure}[ht]
\centerline{\parbox[b]{3in}
{\includegraphics[width=3in,height=2.2in]{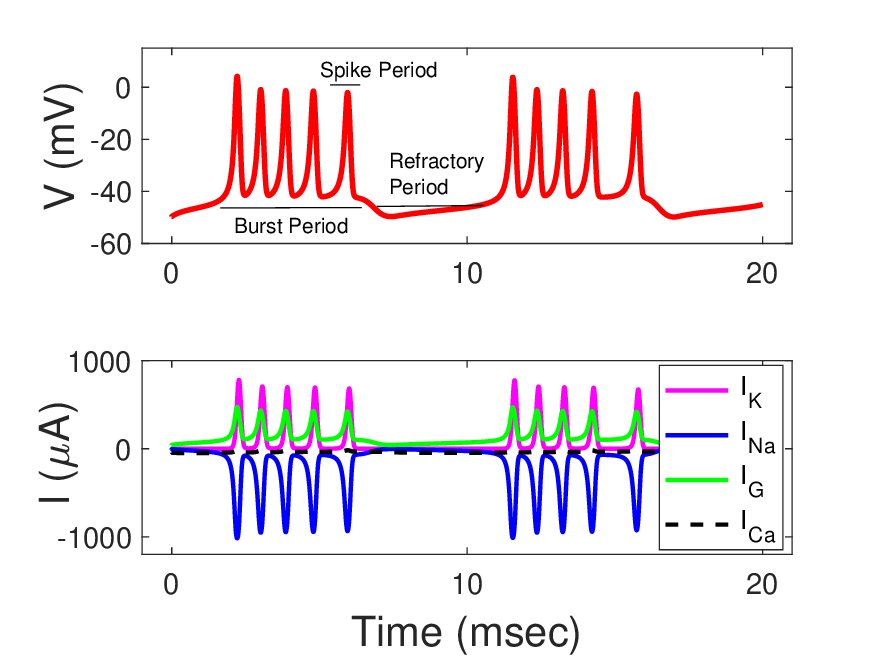}}
\parbox[b]{3in}
{\includegraphics[width=3in,height=2.2in]{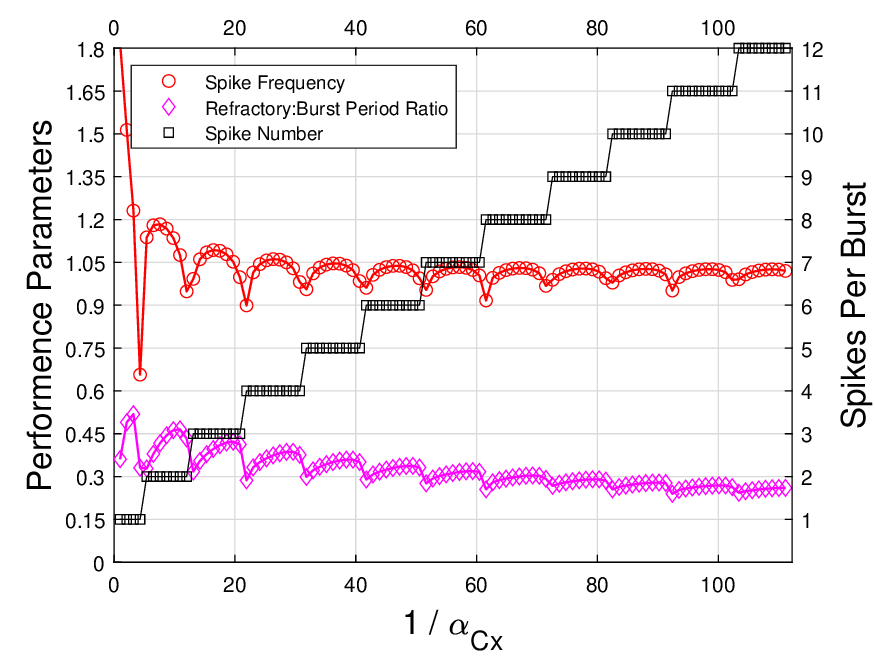}}
} \centerline{(a)\hskip 3in (b)}

\caption{(a) Spike bursts and terminology Legend. (b)
Parameter values for the neural model Eq.(\ref{eqCRmodel}): $\sigma=-1$,
$\Kabattery=-60.0$ mV,
$\Kbarconductance=70$ m.mho/cm$^2$,
$\Kzero=-43.0$ mV, $\Keta=0.04$/mV,
$\Naabattery=45.0$ mV, $\Nabarconductance=332.0$ m.mho/cm$^2$, $\Nazero=-52.0$ mV,
$\Naeta=0.01$/mV, $\Caabattery=40.0$ mV, $\Cabarconductance=20.0$ m.mho/cm$^2$,
$\Cazero=-50.0$ mV, $\Caeta=0.06$/mV, $\Gabattery=-55.0$ mV,
$\Gbarconductance=8.0$ m.mho/cm$^2$, $\Gzero=75$ mV, $\Geta=0.03$/mV,
$C=1\mu{\rm F/cm}^2$, $\KTimeConstant=7.0$/ms, and
$\Kspopen=\Naspopen=\Caspopen=\Gspopen=10^{-5}$. The changing parameter
is for $1/\CaTimeConstant\in [1, 110]$ in ms.
Quantitatively similar result also holds (not shown) for the same parameters
except for $\sigma=+1$, $\Cazero=-42.0$ mV, $\Caeta=1.0$/mV, and
$\CaTimeConstant$ between $0.01$/ms and $1$/ms. All spike bursts
start at the same initial values $\Vvoltage(0)=-49$, $n(0)=m(0)=0$,
$h(0)=1$, and $c(0)=0.025$.}\label{figPlasticSpikeBursts}
\end{figure}
%%%%%%%%%%%%%%%%%%%%%%%%%%%%%%%%%%%%%%%%%%%%%%%%%%%%%%%%%%%%%
%%%%%%%%%%%%%%%%%%%%%%%%%%%%%%%%%%%%%%%%%%%%%%%%%%%%%%%%%%%%%
%%%%%%%%%%%%%%%%%%%%%%%%%%%%%%%%%%%%%%%%%%%%%%%%%%%%%%%%%%%%%

In Fig.\ref{figPlasticSpikeBursts}, all parameters are fixed from \cite{Deng19}
except for the rate parameter $\CaTimeConstant$ which is used as a bifurcation
parameter. The plot is presented
against $1/\CaTimeConstant$ for a better visibility. From
the graph we can conclude immediately that if we denote the first
bifurcation of $k$ spikes by ${\CaTimeConstant}_{,\ k}$, then the
sequence scales like the Harmonic sequence
\[
{\CaTimeConstant}_{,\ k} \sim 1/k
\]
and its renormalization
\[
({\CaTimeConstant}_{,\ k+1}-{\CaTimeConstant}_{,\ k})/({\CaTimeConstant}_{,\ k}
-{\CaTimeConstant}_{,\ k-1})\to 1\hbox{ as $k\to\infty$}
\]
converges to a universal number which is the first natural number 1,
according to the neural spike renormalization theory of \cite{Deng11a, Deng11b}.
Second, the refractory time for the $k$-spike burst is proportional
to the bursting time because their ratio is approximately a
constant around $0.5$, so that the $k$th spike base time
$\tau_k$ for letter $b_k$ is approximately $(1+0.5)\times${\it bursting
time} for $b_k$. Thirdly, from the spike frequency plot
we can conclude that it is approximately a constant around
1 cycle per msec for all spike bursts. As a result,
the $k$-spike burst takes about $k$ msec. All these values
can be obtained by choosing an $\CaTimeConstant$ value from
the $k$-spike interval which is called the isospike interval.
Hence, we can conclude empirically that
\begin{equation}\label{eqtau_ast}
\tau_k=\tau_\ast\times k
\end{equation}
for some constant $\tau_\ast$ around 1.5 msec. Hence,
the hypothesis is satisfied for the information
distribution equation (\ref{eqGenGold}) for the
Golden Ratio distribution ($n=2$) and for
the generalized Golden Ratio distribution ($n>2$).

\begin{figure}
\centerline{\parbox[b]{3.2in}
{\includegraphics[width=3.2in,height=2.4in]{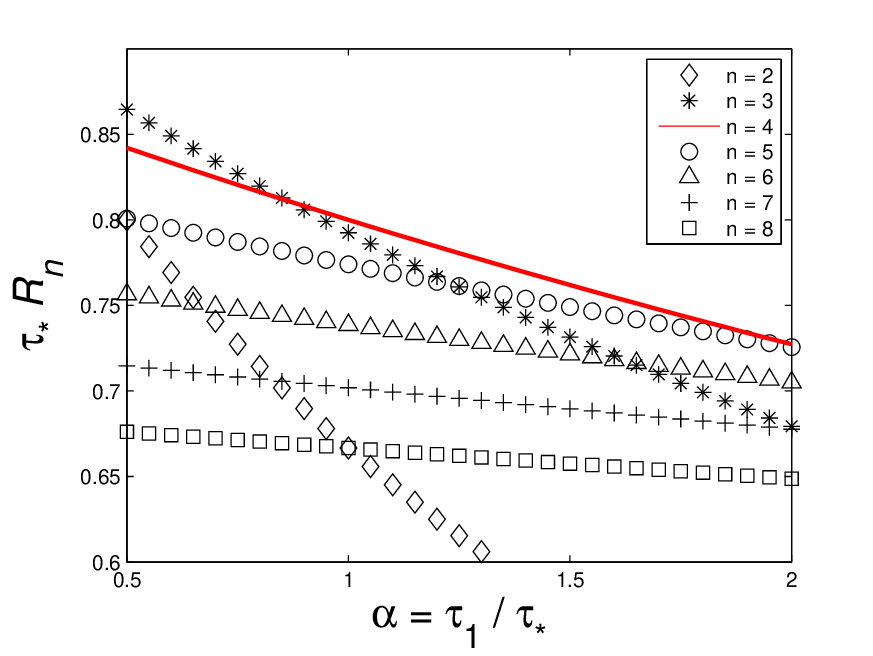}}
\parbox[b]{3.2in}
{\includegraphics[width=3.2in,height=2.4in]{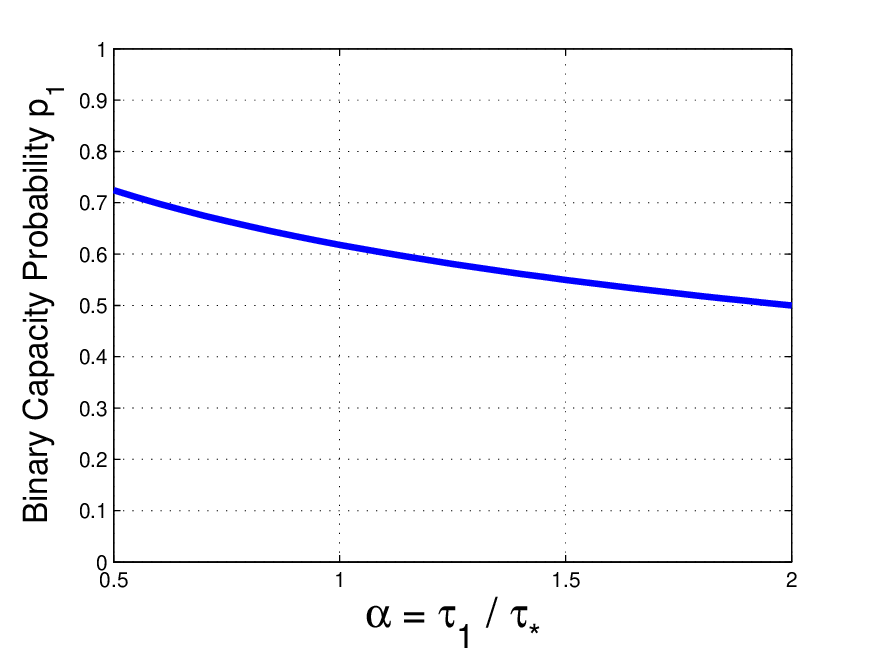}}
} \centerline{(a)\hskip 3in (b)}
\caption{%\leftskip .3in \rightskip .3in
(a) Mean Rate Comparison. $R_4$ seems to be the likely optimal
solution for most practical choices of $\alpha$. (b) Channel
capacity probability $p_1$ for a binary information source.
\label{figQuadrarySEEDCom}}
\end{figure}

\bigskip
\noindent\textbf{5. Discussion.} From Fig.\ref{figPlasticSpikeBursts}
we can see that the 1-spike burst is different from the rest. Although
we can find parameter value from the 1-spike parameter interval to
behave similarly to the rest of the spike bursts in the base
process time $\tau_1$, but in practice, such a value can be hard
to fixed. Instead, its spike-frequency can be higher or lower than
the average of the rest. This is due to a phenomenon, referred
to as spike frequency adaptation (\cite{Guck97, Erme10}), for neural models.
Let $\tau_\ast$ be the same parameter as in (\ref{eqtau_ast}) that
is the average spike-burst period for $k\ge 2$ and
$\tau_k=\tau_\ast k$ for $k\ge 2$. For the
1-spike burst, we express its processing time as a scalar multiple
of $\tau_\ast$ as $\tau_1=\alpha \tau_\ast$ for some $\alpha$
either greater, or equal to, or smaller than 1. Thus,
for the MTR, $R_n$, for the $\mathcal A_n$ code,
$T_n=\sum_{k=1}^n\tau_k/n=\tau_\ast((\alpha-1)/{n}+(n+1)/{2})$
and $R_n$ is
\[
R_n(\tau)={H_n}/{T_n(\tau)}={(\log_2
n)}/[\tau_\ast((\alpha-1)/{n}+(n+1)/{2})].
\]
Fig.\ref{figQuadrarySEEDCom}(a) plots its graph against the parameter
$\alpha$. It shows for the range of $0.5<\alpha<2$,
the optimal MRT is with $\mathcal A_3$ or $\mathcal A_4$. For
the binary alphabet's channel capacity, the optimal distribution
$p_1$ for the 1-spike burst is computed numerically
from the equation (\ref{eqCCapacity}):
\begin{equation}\label{eqApproxGold}
p_1+p_1^{\tau_2/\tau_1}=p_1+p_1^{2/\alpha}=1
\end{equation}
for each $\alpha$ from the interval $[1/110,1]$.
Fig.\ref{figQuadrarySEEDCom}(b) shows the graph of the solution $p_1$
as a function of $\alpha$. It goes through the Golden Ratio
at $\alpha=1$ as expected.

%%%%%%%%%%%%%%%%%%%%%%%%%%%%%%%%%%%%%%%%%%%%%%%%%%%%%%%%%%%%%
%%%%%%%%%%%%%%%%%%%%%%%%%%%%%%%%%%%%%%%%%%%%%%%%%%%%%%%%%%%%%
%%%%%%%%%%%%%%%%%%%%%%%%%%%%%%%%%%%%%%%%%%%%%%%%%%%%%%%%%%%%%
%%%%%%%%%%%%%%%%%%%%%%%%%%%%%%%%%%%%%%%%%%%%%%%%%%%%%%%%%%%%%
%%%%%%%%%%%%%%%%%%%%%%%%%%%%%%%%%%%%%%%%%%%%%%%%%%%%%%%%%%%%%
%%%%%%%%%%%%%%%%%%%%%%%%%%%%%%%%%%%%%%%%%%%%%%%%%%%%%%%%%%%%%

Obviously, nature has build a communication system out of neurons.
The results above suggest that such systems come with
inherent preferences in information. There are may examples
of memory related animal systems where the number of
preferred modes for operation is around 4, the so-called
magic number 4 phenomena,  c.f.
\cite{Mill56,Boyn79,Poll89,Cowa01}. The underlining hypothesis
is that animal brains have a neurological tendency to
maximize information entropy against time in cost.

Information sources with the Golden Ratio distribution are many.
One example is the Golden Sequence, 101101011011010110101...,
which is generated by starting with a symbol 1 and iterating the
sequence according to 2 rules: replace every symbol 1 in the
sequence by 10 and replace every symbol 0 in the sequence by 1.
The distribution $\{p_0, p_1\}$ of the symbols $\{0, 1\}$ along
the sequence has the Golden Ratio distribution: $p_0+p_1=1,
p_0/p_1=p_1/1=\Phi$. This gives a perfect illustration of
Shannon's Fundamental Theory of Noiseless Channel: assigning
symbol 1 to the 1-spike burst base and symbol 0 to the 2-spike
burst base for the source-to-channel encoding gives rise to the
binary neural system's channel capacity. Penrose's aperiodic tiling
is another example with Golden Ratio distribution. In its simplest
form, its bases consist of a 54-degree rhombus and a 72-degree
rhombus. The frequencies with which the rhombi appear in the plane
follow the Golden Ratio distribution (\cite{deBr81}). Again, a
trivial but most natural source-to-channel encoding gives rise to
the fastest source transmission rate for the binary neural code.

Equation (\ref{eqApproxGold}) for $\alpha$ gives a way to quantify
how close something is to the Golden Ratio. For example,
a rectangular frame is uniquely defined by its height-to-width
(aspect) ratio. A frame of the Golden Ratio is
1:1.6180=1:$1/\Phi$. To translate it into a statistical
distribution over a binary source, both height and width need to
be proportionated against the height-width sum. Thus, the
width:sum fraction is $p_1=1/\Phi/(1+1/\Phi)=1/(\Phi+1)=\Phi$, the
Golden Ratio, which in turn corresponds to $\alpha=1$ if
$\{p_1,1-p_1\}$ is the binary channel capacity distribution. The
aspect ratio of a typical wide-screen monitor is 10:16,
corresponding to a $p_1=16/26=0.6154$ binary distribution, and
$\alpha\sim 0.99$ for the neural binary channel capacity. The aspect
ratio of a high definition TV is 9:16, corresponding to a
$p_1=16/25=0.64$ binary distribution, or $\alpha\sim 0.90$. All
fall inside the $R_4$-optimal mean rate range and near the Golden
Ratio distribution for binary channel capacities as shown in
Fig.\ref{figQuadrarySEEDCom}.

In conclusion, with the assumption that the number of spikes 
per burst forms a letter of an alphabet and the processing 
times for the letters progress like the natural number, 
then the neural spike code with the first four letters achieves
the best average information rate, and for any binary source, 
the information distribution in Golden Ratio achieves the 
binary channel capacity for the neural spike code. 
Altogether, our result seems to support
the hypothesis that human's brain is biologically
build for informational preferences.

\bigskip\noindent
\textbf{\large Declarations}

\small\noindent
\textbf{Ethical approval:} Not Applied.

\small\noindent
\textbf{Competing interests:} None.

\small\noindent
\textbf{Authors' contributions:} Not Applied.

\small\noindent
\textbf{Funding:} None.

\small\noindent
\textbf{Availability of data and materials:} All data generated or analysed
during this study are included in this published article. Matlab mfiles
are available from the corresponding author on reasonable request.

\end{document}